# A model of Cross Language Retrieval for IT domain papers through a map of ACM Computing Classification System

Gérald Kembellec, Imad Saleh
Paragraphe Laboratory
Paris 8 University.
Saint-Denis, France
gerald.kembellec@univ-paris8.fr
imad.saleh@univ-paris8.fr

Catherine Sauvaget
LIASD
Paris 8 University.
Saint-Denis, France
cat@ai.univ-paris8.fr

*Abstract*— **This article presents a concept model, and the associated tool to help advanced learners to find adapted bibliography. The purpose is the use of an IT representation as educational research software for newcomers in research. We use an ontology based on the ACM's Computing Classification System in order to find scientific articles directly related to the new researcher's domain without any formal request. An ontology translation in French is automatically proposed and can be based on Web 2.0 enhanced by a community of users. A visualization and navigation model is proposed to make it more accessible and examples are given to show the interface of our tool: Ontology Navigator.**

## I. INTRODUCTION

The purpose of this article is to develop a method for approaching the IT field for the use of student researchers, typically in their second year of master or beginning their PhD thesis. In the French university context, it is common to see students in 2$^{nd}$ and 3$^{rd}$ cycles experiencing real difficulties in collecting the documentation on their field of study or research. In our view, the main idea is to support these learners by a tool to supplement their perception of the knowledge domain "stored" in ontology. It is to consider, even to hope, that mastering tool lets it become obsolete for the advanced learner (became autonomous). Our context is bibliographical research in the IT area for "Insiders" but not experts, who are increasingly lost in a predominantly English corpus. Once the subject and the target audience have been defined, it is necessary to determine the habits, practices, behaviour and attitudes of young researchers in their research of information. An important point of transition between the first and the second year of master is the increased focus on access to quality and credit worthy information. The purpose of information retrieval systems, such as conventional search engines or documentaries on the Internet is to search by keywords or natural language. During the first part of their studies, students tend to seek information on the total mass of the Internet without particular method, discernment or qualitative discrimination. When transitioning to a higher level of studies the method of research evolves. This is mainly due to the fact that access to a huge amount of information causes two major problems. The first major obstacle to efficient research is the difficulty to control the quality of information. The ability to assess the creditworthiness of the information will depend on the knowledge level of the student researcher. The second problem identified in a classic request on a search engine is the amount of information returned by each search. Our proposal, while not purporting to replace the role of a director of research, seeks to lead the researcher in his approach of the field of IT research. In this article, we define the concept of domain ontology and the way it can help to search scientific information. We then describe step by step the design of our tool how it includes translation tools, and discuss the methods of representation. We continue the article by presenting results.

## II. SEARCH FOR INFORMATION BY DOMAIN ONTOLOGIES

By domain ontology we mean a conceptual hierarchy designed by an expert within a structure, where elements are linked by their proximity in terms of syntactic or semantic relations. The traditional approach to the use of ontology's area is to prioritize subsets of the area for management purposes. The ontology is then used most often to prioritize and rank the components of the domain and to describe their relationship. A frequent application is indexing a specialized corpus. A more innovative use of ontology is to reverse the process. It is possible to use domain ontology as a means of research in a text, a corpus, a digital library, or perhaps even the Internet. Thanks to a combination of different semantic technologies, Stephan Bloehdorn has proposed an interesting method of searching in digital libraries [4]. He defined an approach by analysis of structured questions in natural language with a formal grammar. It is then the role of the system to understand the question, identify keywords, titles and authors. Basic examples of questions would be: who wrote this book? What book deals with this specific topic? Which article is part of this conference and corresponds to these keywords? This approach translates natural language into meta-data, and rephrases the question in SPARQL language [12]. As the answer is contained in a Resource Description Framework (RDF) file [2], this enables updates in real time, the use of a variety of formats as well as sourcing from many different locations. This method enables the user to avoid using any database in the common sense of the term.

*A. Concept of user relevance*

IT is a very broad area, which includes a multitude of sub disciplines, and is a powerful tool used in many scientific fields. It is therefore necessary to understand the nature and context of the user's research and his angle of research as much as possible. For example: the couple of words "data storage" will not have the same meaning for an assembly technician, a systems and networks engineer or a librarian. The technician's perception of "data storage" is the hard disk or USB drive. The systems and networks engineer will have a broader vision of "data storage" including not only the concept of devices, but also methods of storage such as NAS, data redundancy (RAID level), information sharing techniques (NetBIOS, NFS, SMB ...) and permission(s) (reading, writing and execution). Finally, the librarian will understand the term "data storage" primarily as an integrated library system, which administers the loans and reservations and manages the order tracking and state of the inventory. These three professionals, having advanced knowledge in their own particular field have different uses of the term "data storage". This is however not a case of polysemy (multiple meanings) but is rather a difference in the angle of perception of these three professionals. The issue of user relevance arises in the particular case of the IRS. The idea of user relevance has greatly influenced the tool, which focuses on the angle of perception of the user and not only on the data.

*B. Maintaining the Integrity of the Specifications*

The template is used to format your paper and style the text. All margins, column widths, line spaces, and text fonts are prescribed; please do not alter them. You may note peculiarities. For example, the head margin in this template measures proportionately more than is customary. This measurement and others are deliberate, using specifications that anticipate your paper as one part of the entire proceedings, and not as an independent document. Please do not revise any of the current designations.

### III. IT FIELD ONTOLOGY: DESIGN OF A USABLE MODEL

Initially, the approach is onomasiological or top-down, i.e. the corpus is classified in a structure which is a finished standardized set. In a second step we purport to enrich the structure, where this becomes necessary, through the adjunction of additional corpus. The domain ontology consists of a tree of topics ranging from a generic root (in this case computer science) to the leaves of knowledge. The arcs are links between nodes which materialize top-down or bottom-up relations or ties of similarity. The ontology contains no articles, but nodes with labels containing keywords issued by superior nodes. These keywords can generate a request to be submitted to the main scientific online libraries.

*A. Proposal of a model*

This project must consist in a tool of flexible use which integrates the field of a particular user to be within the user's grasp. Therefore it should help the user mastering his field of expertise. This tree can simply be seen as the external skeleton or *exoskeleton* in the IT field. First keywords of each node or leaf are the words constituting its label. These keywords are called "native" keywords, as opposed to other keywords added afterwards, which is referred to as "added" keywords. Our starting point was the description of research with a minimal ontological exoskeleton. To put into place such a minimal ontological exoskeleton it was necessary to find taxonomic approaches representing as carefully and as fully as possible the broad field of IT. Then, to conceptualize this field, it was necessary to segment the titles of each branch. This specification phase passes through a stage of construction of keyword "clusters" related to each branch, thanks to lemmas (canonical form of a lexeme) extracted from titles. From a technical point of view, for greater ease of handling, it would be appropriate to integrate the ontology and its keywords in a database, which results in a comprehensive ontology in Extensible Mark-up Language (XML [5]) where developments are updated in real time. For our test phase, the corpus of research is composed of the titles of articles published since 1945 and referenced in the Database systems and Logic Programming (DBLP [10])by Michael Ley from the German University of Trier. It is the source of an XML document of about one million admissions in BibTeX[1] format (format of bibliographic description of $L^AT_EX$[2]). It should be noted that the papers are written in various languages (cf: Section English to French translation). We also propose Meta queries to online digital libraries such as Computer Science Bibliography (CSBIB[3]) or ACM.

*B. Choosing the best reference for IT classification*

We tried as a first step to find an agency specialized in computer sciences. Then we proposed a system of representation in the field that we wish to model. For the sake of simplicity let's take the on-line encyclopedia Wikipedia as a first step. Indeed Wikipedia from an IT perspective is classified according to an internal hierarchy, has an abundant corpus and is immediately available in XML and RDF. Unfortunately, as of today the scientific legitimacy of Wikipedia is not demonstrable. Let us then turn to Computing Classification System [8] (CCS[4]), whose legitimacy is evident. Moreover, conveniently the Association for Computing Machinery (ACM) has its own digital library of scientific articles indexed according to the CCS model. However, for our purposes, the CCS is not usable as it stands. The CCS is more in the state of taxonomy than ontology. According to Grüber, an important aspect of ontology (in addition to clarity, consistency, minimal commitment, and deformation) is scalability [7].

---

[1] http://www.bibtex.org/
[2] http://www.latex-project.org
[3] http://liinwww.ira.uka.de/bibliography
[4] http://www.acm.org/class/1998

*C. The generation of keywords and the emergence of semantic closeness*

Consider the corpus including a mass of titles composing IT domain ontology in terms of the science of measuring information and statistics. According to Le Coadic [9], when we consider a series of scientific articles, we must pay attention to significant words and their co occurrence to generate significant semantic proximities. So when a couple of associated words appear simultaneously in several node labels, it is likely that the subjects in question are associated. Of course, in this case we will use this approach only on the titles but ACM labels appear sufficiently precise to be representative of all articles, both from the general and particular point of view. Thus, the words that best represents the label will be added as keywords of the article and of the branch of the ontology, other words which are less representative of the label will be added as semantically near. Subsequently, during indexing phase of a digital library, if an article appears to be indexed in two places, it is proposed to establish proximity link between the two branches of the ontology.

*D. Corpus Interfaces*

Each library like CSBIB, DBLP, ACM and others has its own scientific query interface. We tried to find the RDF document that describes each database, but it does not exist. It should be noted that if each site provided data Description services such as RDFa [1], this work would be greatly simplified. A database, the "The scientific library of the field of Information Technologies" was created, describing each article by its title, the context and year of publication, and authors of this article. The database, automatically updated each week on an incremental basis, would ideally continuously generate a single RDF document describing the pseudo corpus. The term "continuously" means that in theory for every query, a snapshot of the corpus will be established by RDF through interrogation of the database and will be processed to reflect the weekly updates. The corpus of scientific articles would not be hosted locally on the host machine of the ontology for legal reasons, but also for reasons of storage capacity. This is why we prefer to use the term pseudo-corpus rather than corpus. Indeed labels, and possibly abstracts indexed in digital libraries do not strictly constitute a corpus.

*E. The perspectives for IT ontology*

Let's imagine that we index a corpus composed of titles of articles. Most of the times the titles of scientific articles are long enough to provide a number of keywords indicating the leading ideas. During the phase of indexing the corpus, if an article's title appears as "unclassifiable", we propose to classify it momentarily in a branch of the ontology having the closest semantic proximity within a "miscellaneous" or "general" subsection. Then once the ontology has reached a sufficient size, the "orphan" article will be classified permanently by creating a new branch on the ontology where semantic proximity is the greatest (using added keywords). The process described above is one of the vectors of the evolution of an ontology which is not static but evolves with the corpus and the work of the users and experts. The extensions that may be added to the ontology must be anticipated during its creation. It should be possible to add new concepts without having to modify the foundation of ontology. For example a "branch" which turns up an important number of common keywords would constitute a suitable root for the ontology.

*F. English to French translation*

According to the open letter by a few thousand French researchers to the French Agency for Evaluation of Research and Higher Education (AERES[5]), it is widely recognized that the lingua franca of scientific research today is English. Each researcher should in theory feel comfortable with this international scientific language. Why should we make the effort to translate the titles of the IT ontology branches in French while the body is predominantly English, the predominant scientific language? However even if the user feels comfortable reading technical and scientific texts (as the case may be with a good dictionary in hand), he may feel more at ease in French to conduct his research. We decided to use a Web 2.0 approach, i.e. hybrid translation starting with automatic translation which is thereafter corrected and completed through communal manual translation. The simplest and most economical solution to automate an Anglo-French translation would be to use an online translation tool. The tools that have caught our attention are Babel fish, Yahoo and Google Translate. We designed and used an API wrapping to generate a French version of the ontology based on one of these tools. It can be pointed out that this kind of online applications would benefit from having its own official API. Of course nothing can replace manual translation, which is why we incorporate a notion of folksonomy with a RDF Site Summary (RSS [11]) in the tool. This enables the last user to report a translation error, or imprecision, to the management committee. This group will consist of researchers from laboratories of the research and training unit which will validate the proposal or reject it. According to Thomas Vander Wal, the value of external marking of the folksonomy comes from the users using their own words which add an explicit dimension, which will be an inference of the object [4]. The system of translation of the ontology's nodes automated in a first step continued and developed by English language users and validated by experts, can be carried out without recourse to a professional translator, or occupying an expert on a full-time basis. This procedure implies considerable time saving for researchers and the financial economy should not be underestimated. The technical aspects of this process should be simplified, as much as possible, for the user so as not to discourage him from making a proposal e.g. making a proposal should also not take him more than a few seconds. Once the proposal has been submitted (cf: Figure 1), a RSS feed is generated and remains active until at least two intuitive navigation. In our context, the tool of representation must abide by a number of rules set out by Christophe Tricot and Christophe Roche [13].

---

[5]http://petition.hermespublishing.com/

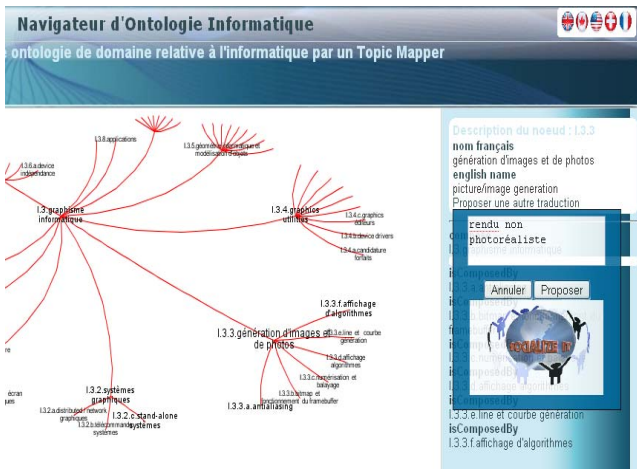

Figure 1. Alternative translation of a node

To be effective a visualization system should observe the following rules as a minimal requirement:

- Provision of an overview of the ontology. This allows the user to identify all the concepts in the field.

- Use of a "focus + context" to allow the user to concentrate on certain aspects while having access to others;

- Use of plane geometry, to avoid disturbing natural perception of the manipulation taking place in the plane. This particular point, has however not been followed in the present case, because giving the mass of data to display and the wish to comply with the other principles, it is complex, if not impossible, to combine a tree display and Euclidean geometry.

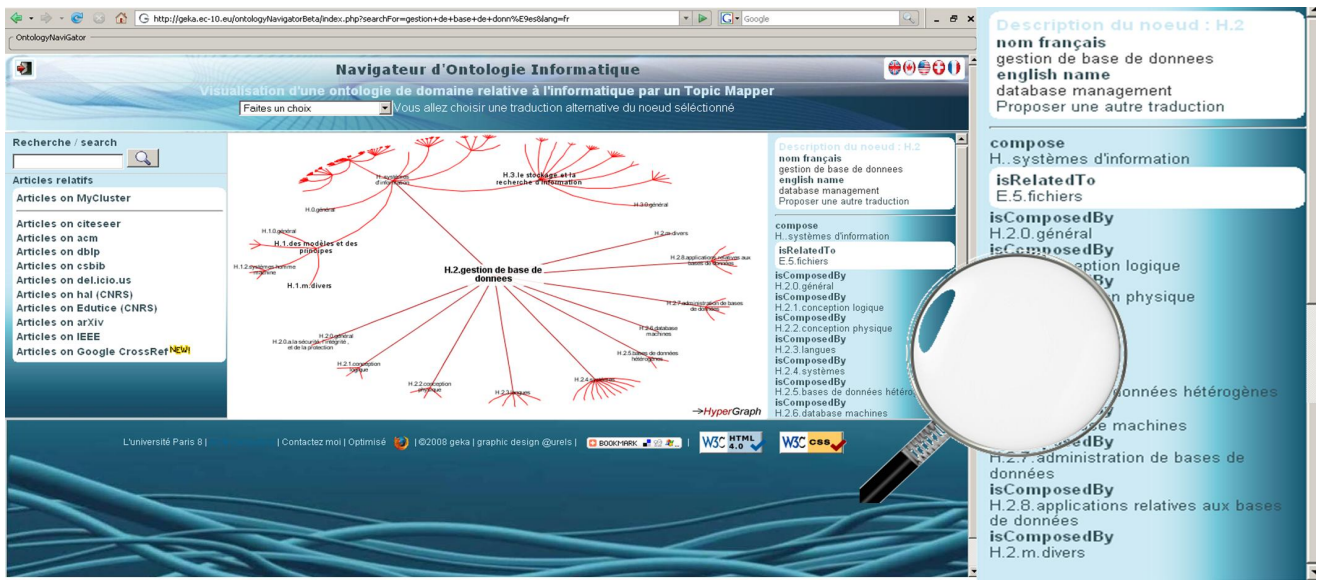

Figure 2. Search scientific articles by navigating through ontology and zoom on the focused context

committee members have verified the proposal. It is contemplated to correct the French part of the ontology over a period of time which is yet to be determined. Another advantage of this process is that it takes into account terminology modification which is inherent to the field of IT. Due to the interaction between the system and the user, the user enriches his knowledge in the field in question while participating in its evolution.

*G. Choosing a model of representation*

To make the corpus more accessible, we propose to facilitate the representation of the domain ontology by representing it as a navigable map. The tree should enable the user to focus on the branch containing a formalization of the concept sought. There are many ways to view ontologies, but all are not specific to navigation, at least not as concerns

According to feedback C. Tricot obtained from an experiment, two types of users emerge: "newcomers" and "experts". Newcomers understand the field and its concepts without perceiving details of the organization and interactions. Experts have a perfect mastery of the entire field both in terms of the content of the concepts and the links that bind them together. For our purposes, users have a profile of a master student or a PhD beginner who searches scientific information on a subject in a specialized field. We tried to find a compromise on the representation of the field offering direct access to context on the element in focus. In the article by C. Tricot, it appears that the model representation by radial tree is the most suitable for experts and the model representation by eye tree is the most suitable for newcomers. The eye tree visualization allows a global view of the field and the possibility of a wide-angle focused (fisheye polar) on a point of detail around which the field is articulated. The main shortcoming of this alternative, in the context, is to be limited to a plane. This prevents putting elements into perspective which is possible with the use of cone trees. The radial tree is quite similar to the eye tree combining global vision of the field and the polar fisheye. But the background and focus is more significant within the graph. It appears however that the very advantages of the radial tree (focus + context) also cause a loss of contact with the primary objective which is to keep the global view. In addition, a radial tree describing the ACM would be quite unreadable because of the huge size of the ontology. In view of the size issue, a visualization of information clusters seems to emerge through the combination of ontology and a technology called Topic Map, thanks to the open source applet Hypergraph. While not specifically conceived to effectively represent ontology, the Topic Map is a hyperbolic tree type representation which consists of mapping the ontology and unlimited navigation. We adapted it to enable angles of perception to stand out as well as their focus and contexts. This method is a hybrid approach between the eye tree and the hyperbolic tree.

## IV. GENERATING A BROWSED META REQUEST SYSTEM

### A. Meta request concept

Through the Topic Map described, we want to provide the advanced learner access to scientific documents relating to his field of research. We intend to use external resources for our application, such as online Knowledge Base System (KBS). For this purpose we define user's context and profile to enable personal customized access to knowledge, through this application, in a transparent manner. This is a Reverse-Engineering approach of interrogation of the external KBS. We call Meta query, a query sent to a remote KBS without knowing the system of internal questioning. This is done by simulating a manual use of the remote application through combination of lemmas of keywords extracted from the context of navigation.

### B. Modelling the system

While a natural language search on all words in the order established has little chance of success, a search by key words has every chance to return hundreds of thousands of results. The first step in generating request is the filtering of "noise" on the label when positioning the user's browser in the ontology thanks to the stop-lists (one in each language), which eliminates empty words, like pronouns and nouns which are too common for significant meaning. A similar preliminary stage is conducted when using a search engine in a natural language search. The second step is the lemmatization[6] of words, followed by a calculation of statistical proximity of all of the words which have emerged from the keyword cluster in a branch of the ontology. It may be appropriate to provide a valuation of the keywords in this context? This point could be the subject of a further study.

---

[6]Lemmatization is the process of finding the normalized form of a word.

### C. Trial of navigated search and results

The first stage of the research is to navigate down the tree until the node that is the most representative of the concept sought. The context block (cf: Figure 1) offers a direct access to online digital library articles as CSBIB, DBLP, or ACM by generating contextual meta-queries to these sites. We call these queries meta-queries because they do not directly generate a request, but an URL with keywords. The remote Knowledge Base System (KBS) uses its own search engine to generate the real request. But the tool also proposes to search the internal database of titles of articles. In the example, search for "database management" is generated and proposes several dozen results. We choose the article: "Managing taxonomies in relational databases". The database provides us with the name of the principal author. The tool checks for the presence of an URI on the article in the database. In the absence of an URI a request to Google Scholar is automatically generated, which provides us with a direct access to the article (cf: Figure 3). Tests were performed on the classical databases, but results pertaining exactly to the subject of research are still too few. This mechanism for generating requests is still in a heuristic stage, but opens interesting prospects.

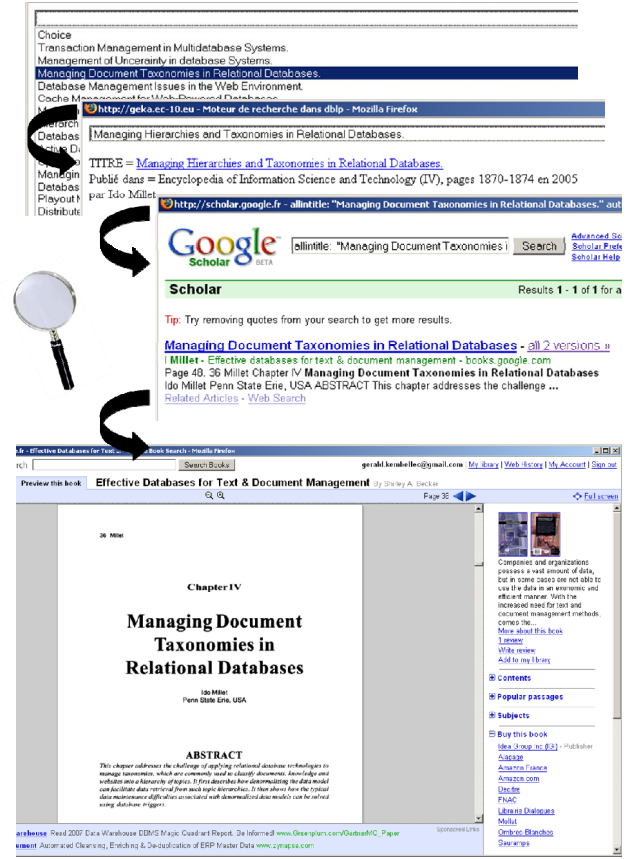

Figure 3 Scientific articles proposed by the system

Our tool is accessible online, but the fee paid by the University's Library to ACM portal only grants access within the University. Nevertheless, the tool also uses several free online databases. A feedback form is available to get feelings and comments about the tool. We used the answers to measure final user's interest. The questions were asked about the usability, the intuitiveness and the results produced by the tool. We also measured users' habits with cookies and Google Analytics. The most often comprehensive problem that users had with the tool was the lack of intuitiveness. The cross language search based on Meta data gave some lack of outcome as described in the following paragraph.

We experienced our tool within our University (Paris 8 in France) in Computer Sciences department, in the Library Sciences one, and with a group of 15 Master degree students in "Technology of Hypermedia".

TABLE I. RESULTS

| CCS node | Cross Language Browsing Tests In French | | |
|---|---|---|---|
| | *English Label* | *French Query* | *% of relevance* |
| H.3 | information storage and retrieval | le stockage et la recherche d'information | 100 (cf : Figure 4) |
| G.3 | probability and statistics | probabilités et statistiques | 90 |
| J.6 | computer-aided engineering | ingénierie assistée par ordinateur | 80 |
| F.1 | computation by abstract devices | calcul sur système virtuel | 100 |
| F.2.1 | numerical algorithms and problems | algorithmes numériques et problèmes | 90 |
| B.5 | register-transfer-level implementation | mise en œuvre d'un niveau de registre de transfert | 70 |
| B.5.2 | design aids | aide à la modélisation | 20 |
| B.3 | memory structures | structures de mémoire | 100 |
| I.2.7 | natural language processing | traitement du langage naturel | 100 |
| C.5 | computer system implementation | implantation de systèmes informatiques | 80 |
| D.3.2 | language classifications | classification des langages | 60 |
| E.1 | data structures | structures de données | 100 |
| K.2 | history of computing | histoire de l'informatique | 20 |
| A.2 | reference (e.g., dictionaries, encyclopedias, glossaries) | référence (par exemple, dictionnaires, encyclopédie et glossaires) | 0 |
| I.3.3 | picture/image generation | génération d'images et de photos | 60 |
| | | Medium relevance | 71 |

a.)limited at ten first results

A PhD candidate in computer graphics has done research in ontologyNavigator with the French sentence "*rendu non photo réaliste*" (Non-Photorealistic Rendering). Research in ontology failed. The tool printed: "*rendu non photo réaliste does not exist in French in the ACM ontology*" what we knew to be false in English. The articles on the NPR are usually classified under I.3 and I.4 nodes of the ACM Computer Classification System. These two nodes are respectively labeled "*Computer Graphics* and *image processing*" and "*Computer vision*". We then tried to use the folksonomy option (cf: Figure 1) for the automated processing of language, to propose an alternative translation for the ontology node label. For example, the node I.3.3 "*generation of images and photos*" originally "*picture / image generation*" was given the alternative French proposal "*rendu non-photoréaliste*" (NPR). This proposal has no chance of being selected as the best translation by the committee of experts. This is in fact not a real translation of the node label. It is a specification and not equivalence. However, this proposal gave a result in the next time query because it created a specific entry for the French sentence "*rendu non-photoréaliste*". This notion, if it is used by researchers, allows users to include concepts of equivalence or specification outside of the simple syntax correction.

This option allows the tool to avoid terminology tendencies of the moment. For instance, in French, "*rendu non-photoréaliste*" (NPR) referred to the previous paragraph is not at the time of writing these lines transcription of the most widely used concept involved. The denial word "non" in "*rendu non-photoréaliste*" conveys a negative image. Because of that fact, french specialists more likely uses "*rendu expressif*" (Expressive rendering) for about two years.

The participatory community (folksonomy) also allows members to correct the shortcomings of automated processing of language. It is certain that the growing number of users of the tool significantly affect the quality of research results. This tool has the flexibility of a virtual index on an evolving corpus and presents a possible match between needs of knowledge and virtual location of online IT scientific articles for the young researcher.

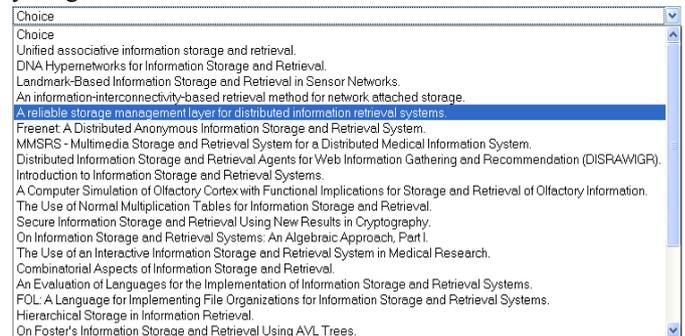

Figure 4. Partial result for a query

V. LIMITS AND PROSPECTS

The testing of the tool by users has shown that the adequacy of current metadata queries generated is relevant. Nevertheless the results are sometime poor or too big on external databases, but more precise on our own database. However, the more the ontology's content is enriched with articles, the more research and indexing becomes accurate. For this purpose, we index a pre-existing corpus of an important size. Another limitation is the physical access to articles which is often subject to the payment of a subscription fee or even a fee per article. That is why it is easier to implement the solution in a university laboratory or a library. However, the use of proxy should help extend access to digital libraries for an entire campus. The tool will be made available to students in their second cycle of studies in the IT department and for the computation centre of the University of Paris 8. We made an online form available for the purpose of recording feedback and follow the evolution of the users. In the near future we plan to extend the application with an ontology based on the Friend of a Friend format (FOaF [6]) for a better understanding of the working groups, teams, and laboratories as well as links to disciplinary transversally. Another goal that we have in mind is to make the system as independent as possible. Possibly the hyperbolic tree / eye tree type navigation system will be modified if another way of displaying the tool emerges. To facilitate the use of items found and selected by young researchers, an interesting feature could be developed in the form of one or several thematic bibliographies on BibTeX format and thus reusable in every article and shareable with researchers with similar profiles.

## REFERENCES

[1] B. Adida and M. Birbeck. *RDFa primer embedding structured data in web pages.* W3C recommendation, W3C, March 17 2008.
http://www.w3.org/TR/2008/WD-xhtml-rdfa-primer-20080317/l.

[2] D. Beckett. Rdf/Xml Syntax Specification. W3C recommendation, W3C, February 10 2004.
http://www.w3.org/TR/rdf-syntax-grammar/.

[3] T. Berners-Lee, R. Fielding, and L. Masinter. World Wide Web Consortium supports the IETF URI standard and iri proposed standard. W3C recommendation, W3C, January 26 2004.
http://www.w3.org/2004/11/uriiripressrelease.html.en

[4] S. Bloehdorn, P. Cimiano, A. Duke, P. Haase, J. Heizmann, I. Thurlow, and J. Völker. *Ontology-based question answering for digital libraries.* In L. Kovács, N. Fuhr, and C. Meghini, editors, *(ECDL 2007), Budapest, Hungary*, Lecture Notes in Computer Science. Springer, Berlin–Heidelberg, Germany, September 16-21 2007.

[5] J. Boyer and G. Marcy. XML 1.1. W3C recommendation, W3C, January 29 2008.
http://www.w3.org/TR/2008/PR-xml-c14n11-20080129.

[6] D. Brickley and L. Miller. *Foaf vocabulary specification 0.91.* Recommendation, RDF and SemWeb developer community, November 2 2007.
http://xmlns.com/foaf/spec/20071002.html.

[7] T. R. Gruber. Toward principles for the design of ontologies used for knowledge sharing? Int. J. Hum.-Comput. Stud., 43(5-6):907–928, 1995.

[8] T. Horton. *Computing Classification System.* ACM description, ACM, July 16 1998.
http://www.acm.org/class/1991/homepage.html.

[9] Y. F. Le-Coadic. Mathématique et statistique en science de l'information et en science de la communication. In IBICT, 2006., 2006.

[10] M. Ley and P. Reuther. Maintaining an online bibliographical database: The problem of data quality. In EGC. Cépaduès-Éditions, 2006.

[11] E. Miller and al. W3C *RSS 1.0 News feed creation how-to.* W3C recommendation, W3C, January 21 2001.
http://www.w3.org/2001/10/glance/doc/howto.

[12] E. Prud'hommeaux and A. Seaborne. Sparql query language for RDF W3C recommendation. W3c recommendation, W3C, January 15 2008.
http://www.w3.org/TR/2008/REC-rdf-sparql-query-20080115/.

[13] C. Tricot and C. Roche. Visualisation of ontology: a focus and context approach. In *InSciT2006*, 2006.

[14] T. V. Wal. Understanding folksonomy (tagging that works). In *dConstruct*, 2006.